\documentclass[12pt]{article}
\usepackage{epsfig}

\makeatletter
\@addtoreset{equation}{section}

\def\baselinestretch{1.2}
\def\href#1#2{#2}
\def\norm#1{ : \! #1  \! :}

\newcommand{\tv}{\tilde{V}}
\newcommand{\hp}{\mbox{${\pi \over 2}$}}

\newcommand{\be}{\begin{equation}}
\newcommand{\ee}{\end{equation}}
\newcommand{\beq}{\begin{eqnarray}}
\newcommand{\eeq}{\end{eqnarray}}

\newcommand{\tr}{{\rm Tr}\,}

\newcommand{\ra}{\rangle}
\newcommand{\la}{\langle}

\begin{document}
\begin{titlepage}

\begin{flushright}
hep-th/0111092\\
\end{flushright}

\vfil\vfil

\begin{center}

{\Large{{\bf  Observables of String Field Theory}}
}

\vfil

\vspace{5mm}

Akikazu Hashimoto  and N. Itzhaki\\

\vspace{10mm}

Institute for  Advanced Study\\ School of Natural Sciences\\
Einstein Drive, Princeton, NJ 08540\\

\vfil

\end{center}

\begin{abstract}
We study gauge invariant operators of open string field theory and
find a precise correspondence with on-shell closed strings. We provide
a detailed proof of the gauge invariance of the operators and a
heuristic interpretation of their correlation functions in terms of
on-shell scattering amplitudes of closed strings.  We also comment on
the implications of these operators to vacuum string field theory.
\end{abstract}

\vspace{0.5in}

\end{titlepage}
\renewcommand{\baselinestretch}{1.05}  

\section{Introduction}

There are two kinds of physically meaningful quantities one can
compute in a gauge theory. One is on-shell scattering amplitude (the
S-matrix), and the other is off-shell correlation functions of gauge
invariant quantities.  In perturbative string theory, however, we only
know how to compute on-shell quantities.  String field theory provides
an interesting opportunity to explore off-shell issues along the line
of standard gauge field theories. In this framework, we have an action
which is invariant under a certain gauge transformation. It is then
natural to raise the questions: what are the operators which are
invariant under this gauge transformation, and what is the meaning of
the correlation functions of these operators?

Different but equally interesting is the question: where are the
closed strings in open string field theory? After all, open strings
are not consistent by themselves since they can self interact to form
a closed string. Indeed, shortly after Witten's paper
\cite{Witten:1986cc} it was shown that at one loop there are poles
which can be related to closed strings \cite{Freedman:1988fr}.  Thus
closed strings exist as a virtual state in the internal propagator of
a generic string diagram.  Unitarity then implies that they should
also appear as asymptotic states, but how does one incorporate closed
strings in a Feynman diagram computation of open string field theory?

One way to introduce closed strings is to include them from the
beginning by considering an open-closed string field theory
\cite{Zwiebach:1991qj,Zwiebach:1998fe}. This is a theory where both
open and closed string fields are defined off-shell.  Such a theory,
however, is difficult to interpret especially in light of new insights
provided by the AdS/CFT correspondence \cite{Maldacena:1998re}.
There, the on-shell observables of closed string theory in the bulk
correspond to off-shell observables of the field theory on the
boundary \cite{Gubser:1998bc,Witten:1998qj}.  It is however very
difficult to see how off-shell closed string observables can have a
dual boundary description.  This implies that string field theory of
closed strings should not exist, at least in an anti de Sitter
background. What the AdS/CFT correspondence really teaches us is that
the two questions in string field theory, one regarding the off-shell
observables and the other regarding the status of closed strings, are
closely related. The goal of this article is to explore this
correspondence.

In fact, similar ideas can be exploited to derive the gauge invariant
open Wilson lines of non-commutative gauge theories from string theory
\cite{Okawa:2000sh,Liu:2001ps}.  There, the way one obtains the
straight open Wilson line of \cite{Ishibashi:1999hs,Gross:2000ba} is
via a summation of all diagrams with one closed string and arbitrary
number of open strings.  In the appropriate decoupling limit, this
infinite sum of higher order disk amplitudes resummed precisely to the
path ordered exponential of gauge fields forming an open Wilson line.

One can imagine repeating the analysis of \cite{Okawa:2000sh} and
resum infinitely many disk diagrams to derive something resembling a
Wilson line of open string field theory.  It is natural to expect that
this will be far more complicated a task for a theory as complicated
as string field theory. Contrary to the expectation, the gauge
invariant combination of open string field which couples naturally to
the closed string is far simpler. We find that a term linear in open
string field derived from a disk amplitude with one open and one
closed string vertices is invariant by itself. Formally, these
operators define a set of Chern-Simons one-forms of open string field
theory.

The set of gauge invariant observables of string field theory
therefore consist of the S-matrix of external open string fields,
correlation function of gauge invariant operators, and combinations
thereof.  Since the gauge invariant operators of string field theory
are in one to one correspondence with the on-shell closed string
vertex operator, these observables have a natural interpretation as
the S-matrix of both open and closed strings. Perhaps a useful point
of view to adopt is that the open strings are the fundamental degrees
of freedom of the theory, and that the closed strings are merely an
artifact of insertions of off-shell gauge invariant operators.

The paper is organized as follows. In section 2, we provide a
schematic construction of the gauge invariant operators and provide a
formal proof that they are gauge invariant.  We also present a formal
relation between the off-shell correlations function of the operators
and the on-shell scattering amplitudes of the closed strings.  In
section 3, we provide a rigorous construction and a proof of gauge
invariance by formulating these operators explicitly in terms of world
sheet oscillators. In section 4, we consider the issue of gauge
invariant observables for the vacuum string field theory.

\section{Formal construction}

In this section we construct the gauge invariant operators, prove that
they are gauge invariant, and interpret their correlation functions
using formal arguments.

\subsection{Gauge invariant operators}

The cubic string field theory is defined by the action
\begin{equation}
S = \int A* QA + {2 \over 3} A*A*A \ ,  \label{action}
\end{equation}
which is invariant under the gauge transformation
\begin{equation}\label{r}
\delta A = Q \Lambda + [A,\Lambda] \  \label{gaugetrans}.
\end{equation}
The fact that the action (\ref{action}) resembles Chern-Simons theory
is a useful hint in constructing gauge invariant operators, since one
can consider other Chern-Simons invariants as long as we ignore the
issue of ghost numbers.  For example, the Chern-Simons one-form
\be\label{l}
\int A,
\ee
is also invariant under (\ref{r}) and can be interpreted formally as
a Wilson line around some one-cycle.

In string field theory, however, one must take the ghost number into
account.  We use the convention that the gauge parameter $\Lambda$
carries ghost number zero, and the physical open string field $A$
carries ghost number one. With that convention, the integral is
non-vanishing only when the integrand has ghost number three. As a
result, the Chern-Simons one-form (\ref{l}) is missing two units of
ghost number and trivially vanishes.

To make proper sense out of 
(\ref{l}), we need to introduce a new ingredient which supplies two
units of ghost number.  A natural candidate is the closed string
vertex operator
\begin{equation}\label{k}
V(\sigma_+,\sigma_-) = c_+(\sigma_+) c_-(\sigma_-) {\cal O},
\end{equation}
where ${\cal O}$ is a conformal primary operator of dimension (1,1) with
ghost number zero.\footnote{The dilaton is a special case where ${\cal
O}$ has ghost number zero but is not a primary conformal field
\cite{Terao:1987ux}.} One can then consider a quantity 
\begin{figure}
\centerline{\psfig{file=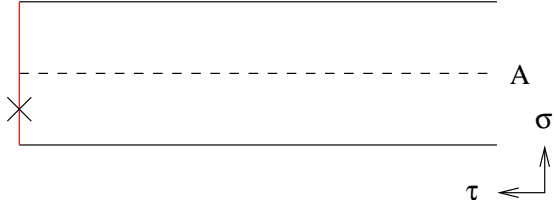}}
\caption{The closed string is inserted at $\tau =0$ at some
$\sigma$. The open string is inserted as usual at $\tau- -\infty$. The
dashed line represents the midpoint.\label{figa}}\end{figure}
\be
O_V = \int V(\sigma_+, \sigma_-) A\ .\label{o}
\ee
From the open string field theory point of view, $V$ is simply an
operator which acts on a string field, much like the BRST or any other
operator.  For each closed string vertex there is such an operator.
Thus eq.(\ref{o}) is a perfectly well defined quantity, in the sense
that it has the correct ghost number and has a concrete oscillator
realization.  One can think of (\ref{o}) as a Chern-Simons one-form
where the role of the ``one cycle'' is played by the closed string
vertex operators. In order to respect the time-ordering in the
operator realization of the conformal field theory, we set $\tau=0$ so
that $\sigma_\pm = \pm i \sigma$. We therefore write the vertex
operator simply as $V(\sigma)$.  In terms of world sheet, this amounts
to inserting a vertex operator at coordinate $\sigma$ at time $\tau=0$
on a state with quantum number of $A$ which has propagated from $\tau
= -\infty$, as illustrated in figure \ref{figa}.

\begin{figure} \centerline{\psfig{file=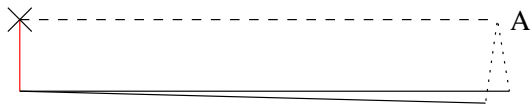}}
\caption{World sheet description of the gauge invariant operator
(\ref{oneform}) in the coordinate where the metric is flat everywhere
but at the insertion of the closed string vertex.  \label{figbb}}
\end{figure}

Let us show that (\ref{o}) with the on-shell closed string vertex
operator inserted at the midpoint $\sigma = \hp$
\begin{equation}
O_V = \int V(\hp) A \label{oneform},
\end{equation}
defines a gauge invariant operator of string field theory.  The
integral identifies the left and the right halves of a string, making
the world sheet take on a geometry of the form illustrated in figure
\ref{figbb}. These operators where first studied by Shapiro
and Thorn \cite{Shapiro:1987gq,Shapiro:1987ac}.

\begin{figure}
\centerline{\psfig{file=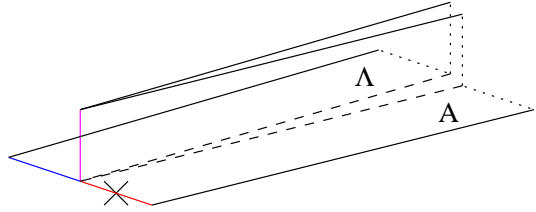}}
\caption{World sheet description of closed string vertex operator $V$
acting on a state which is a product of the form $A*\Lambda$.  The
geometry is such that the metric is flat away from the
midpoint.\label{figb}}
\end{figure}

In order to demonstrate the gauge invariance of (\ref{oneform}), let
us consider the linear and the non-linear contribution to gauge
transformation (\ref{r}) separately.  At the linear
level,\footnote{Gauge invariance of (\ref{oneform}) at the linear
level was demonstrated in \cite{Shapiro:1987gq,Shapiro:1987ac}.}
(\ref{oneform}) transforms according to
\begin{equation}
\delta O_V = \int V(\sigma) Q \Lambda \ =  \int Q V(\sigma)  \Lambda = 0\ , \label{trans1}
\end{equation}
where we used the fact that the closed string vertex operator
$V(\sigma)$ commutes with $Q$ provided that it is on-shell.  Note that
this part of the argument does not depend on $\sigma$.  At the
non-linear level, (\ref{oneform}) transforms as
\begin{equation}\label{1a}
\delta O_V = \int V(\hp) (A * \Lambda) -
V(\hp) (\Lambda * A) \ . \label{nonlinear}
\end{equation} 
To make sense out of this expression, we need to understand what it
means to act with an operator on a string field which is a $*$-product
of a pair of string fields.  In the coordinate where the metric is
flat, the world sheet of the $*$-product of $A$ and $\Lambda$ looks
like figure \ref{figb}.  As long as $\sigma$ is not exactly $\hp$,
there is a neighborhood whose metric is identical to the metric of an
ordinary strip. Therefore, there is a natural coordinate $\sigma_+$
and $\sigma_-$, as well as a meaningful notion of $\partial_+$ and
$\partial_-$.  The midpoint $\sigma=\hp$ appears to be a potentially
singular point in this picture. Let us therefore consider inserting
$V$ at $\sigma = \hp-\epsilon$.  For concreteness, we take $\epsilon$
to be positive.  Then, since the left side of $A*B$ is the left side
of $A$ we get
\begin{equation}
V(\sigma) (A*\Lambda) = (V(\sigma) A) * \Lambda, \qquad \sigma < \hp.
\end{equation}
Thus,  eq.(\ref{1a}) reads
\begin{equation}
\delta O_V = \int (V(\sigma) A) * \Lambda - \int (V(\sigma) \Lambda) * A =
\int  (V(\sigma) A) * \Lambda - \int A *  (V(\sigma) \Lambda)  \ .
\end{equation}
Recalling that the string field theory integral identifies the left
side of the string with the right side of the string, one has
\begin{equation}
\int A* V(\sigma) \Lambda = \int V(\pi - \sigma) A*  \Lambda,
\end{equation}
so that 
\begin{equation}\label{22}
\delta O_V = \int ((V(\sigma)-V(\pi-\sigma))A)*\Lambda  \ , 
\end{equation}
which vanishes for $\sigma = \hp$.  This shows that, much like in
non-Abelian gauge theories, gauge invariance at the non-linear level
imposes non-trivial constraints on objects which are invariant at the
linear level.

We have therefore succeeded in demonstrating that (\ref{oneform}) is
gauge invariant both at the linear and the non-linear level provided
that $V$ is an on-shell closed string vertex operator.  However, this
proof involved several formal manipulations which are quite
subtle. For example, gauge invariance requires that $V$ be inserted at
the midpoint even though the proof of gauge invariance works only if
we approach the midpoint as a limit. Since the geometry of the world
sheet illustrated in figure \ref{figb} is singular precisely at the
midpoint, care must be taken to make sure that the limit
exists. Closely related issue is the fact that operations such as
integration and $*$-multiplication frequently involve insertion of
ghost number at the midpoint. These insertions are often sources of
anomalies whose cancellation is rather delicate.  In order to address
these issues, we will formulate and study the gauge transformations of
(\ref{oneform}) explicitly in terms of oscillators in section
\ref{nb}.

\subsection{Correlation functions}\label{4e}

In the previous subsection, we constructed a set of expressions of the form
\begin{equation} \label{gop}
O_V = \langle  I | V(\hp) | A \rangle,
\end{equation}
where $V(\sigma)$ is the closed string vertex operator
\begin{equation}
V(\sigma) = c_+(\sigma) c_-(\sigma) {\cal O}(\sigma),
\end{equation}
and provided a formal proof of their gauge invariance.  They are
therefore the gauge invariant operators of string field theory. In
this subsection, we will describe the interpretation of the
correlation functions of these operators.

By construction, the operators (\ref{gop}) are in one to one
correspondence with the closed string vertex operators.  This suggests
that the correlation functions of operators of the form (\ref{gop})
should be interpreted as the scattering amplitude of closed strings
through world sheet with boundaries.  Consider for example a two point
function
\be
\la O_1 O_2 \ra =\la \int V_1 A \int V_2 A\ra\ .
\ee
To leading order in perturbation theory, this amplitude is evaluated
by contracting the open string field $A$'s using the propagator. In
Feynman-Siegel gauge, this becomes
\be
\la O_1 O_2 \ra =\la \tilde{I}| {\cal O}_1 |b_0\int_{0}^{\infty} d\tau 
\exp(-\tau L_0) |{\cal O}_2 |\tilde{I}\ra, \label{diaga}
\ee
which can be interpreted as the scattering amplitude of two closed
strings on a disk similar to the ones considered in
\cite{Klebanov:1996ni}. The $\tau$ integral is the integral over the
moduli-space on the world sheet. We see that this space is one
dimensional, appropriate for the amplitude of this type.  It would be
interesting to explicitly reproduce the result of
\cite{Klebanov:1996ni} in full detail starting from (\ref{diaga}).

\begin{figure}
\centerline{\psfig{file=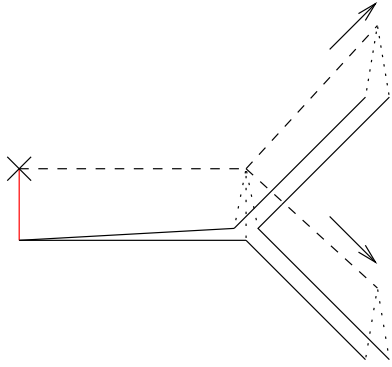}}
\caption{Open string field theory description of one closed two open
string disk amplitude.  The closed string is described by a gauge
invariant operator which propagate to the interaction
point.\label{figd}}
\end{figure}

Another interesting observable to consider is the amplitude of the
type illusterated in figure \ref{figd}, which is obtained by
contracting (\ref{gop}) with the $A^3$ term in the string field theory
action.  This is analogous to the calculation of gluon scattering in
ordinary gauge theories when the action is deformed by gauge invariant
operators, say, $\tr F^4(p)$.  The natural interpretation of this
amplitude is the scattering of one closed and two open strings on a
disk \cite{Hashimoto:1996kf}.  It would again be interesting to
reproduce this result from string field theory. The dimension of
moduli space is certainly in agreement with the expectation from
string field theory.

The most striking feature of the operator (\ref{gop}) is the fact that
it is linear in open string field. This is in marked contrast to the
case of open Wilson lines in non-commutative gauge theories where
terms higher order in non-commutative gauge fields were necessary to
ensure gauge invariance at the non-linear level. The remarkably simple
structure of (\ref{gop}) can be understood in light of the work by
Zwiebach \cite{Zwiebach:1992bw} who showed that a theory with an
action of the form
\begin{equation}
S = \int A * QA + {2 \over 3} A*A*A + \int A \Psi \label{zwbact},
\end{equation}
where $\Psi$ is a closed string field, can be understood as a special
limit of open closed string field theory provided that $\Psi$ is
on-shell. In particular, this theory covers the full moduli-space of
the scattering amplitudes of open and closed strings with a boundary.
The $A\Psi$ term is to be interpreted as (\ref{gop}). The fact that
all such amplitudes can be generated by gluing interaction vertices of
(\ref{zwbact}) through propagators suggests that (\ref{gop}) is the
complete description of the coupling of closed strings.

These arguments appear to suggest that gauge invariant off-shell
observables encode the closed string on-shell physics, very much along
the lines of AdS/CFT correspondence. In other words, the standard
perturbative on-shell scattering amplitudes involving both the open
and the closed strings capture the complete set of the on-shell and
the off-shell observables of string field theory.  This shift in
perspective may seem at first as nothing more than a complicated
reformulation of well known perturbative string physics in terms of
open string field theory.  Nonetheless, this point of view may provide
the critical insight in formulating string theory at the
non-perturbative level.  One important consequence of this new
perspective is the fact that even though open strings are taken
off-shell, the closed strings appear only on-shell.  So far, these
arguments have been presented only for amplitudes involving at least
one boundary.  It would be very interesting to see if viewing the
closed strings as off-shell gauge invariant operators proves to be a
useful point of view in thinking about purely closed string processes
(such as the Virasoro-Shapiro amplitudes) in the framework of open
string field theory.

\section{Explicit oscillator representation}\label{nb}

In the previous section, we provided an attractive picture of a
correspondence between gauge invariant operators of open string field
theory and the closed strings on shell. However, some of the arguments
used in the previous section to construct and prove the gauge
invariance of these operators were somewhat formal and could suffer
from number of technical problems when formulated more
explicitly. There are in fact two potential sources of difficulties
having to do with the fact that the closed string vertex operator is
inserted at the midpoint of the open string world sheet.  One is the
fact that the closed string vertex operator inserts a ghost number at
the midpoint. This is dangerous 
since cancellation of ghost related anomalies at the midpoint is
generally subtle.  The other is the fact that the insertion of closed
string vertex operator at the midpoint of a product of two string
field, as we did in establishing the gauge invariance at the non-linear
level, is subtle. In the previous section, we defined this procedure by
inserting the operator in such a way that it approaches the midpoint
as a limit. This limit could potentially suffer from anomalies.  In
order to address these concerns, it is worthwhile to study these
operators more explicitly.

Explicit computations in string field theory is performed using the
oscillator formalism of \cite{Gross:1987ia,Gross:1987fk}. 
In this formalism, a string field is
an element of the open string Hilbert space
\begin{equation}
A \leftrightarrow |A \rangle,
\end{equation}
obtained by acting on the vacuum with operators in the mode expansion
of the string coordinates
\begin{eqnarray}
X_\pm^\mu (\sigma_\pm) &=& {1 \over 2} x_0^\mu +  {i \over 2} \sqrt{2 \alpha'}
\sum_{m = 1}^\infty \left(
  {1 \over \sqrt{m}} a_m^\mu e^{m(\sigma_\pm)}
- {1 \over \sqrt{m}} a_{-m}^\mu e^{-m(\sigma_\pm)} \right), \nonumber \\
x_0 &=& {i  \over 2} \sqrt{2 \alpha'} (a_0 - a_0^\dagger),  \\
b_\pm(\sigma_\pm) &=&
 \sum_{m=-\infty}^\infty b_m e^{i m (\sigma_\pm)} \ , \nonumber \\
c_\pm(\sigma_\pm) &=&
 \sum_{m=-\infty}^\infty c_m e^{i m (\sigma_\pm)} \ \nonumber . \label{modes}
\end{eqnarray}

Formal operations such as integration, conjugation, and
$*$-multiplication are defined in terms of special states
$|I\rangle$, $|V_2 \rangle$, and $|V_3\rangle$
so that
\begin{eqnarray}
\int A = \langle I|A \rangle \label{int},\;\;\;\;\;
{}_1\langle A | = {}_{12} \langle V_2| A \rangle_2,\;\;\;\;\; 
|A*B \rangle_3 =   {}_1\langle A | \ {}_2 \langle B |V_3 \rangle_{123}.
\end{eqnarray}
The subscripts on the brackets label the open string Hilbert spaces
when more than one are involved.  Explicit expression for these  states
can be found in \cite{Gross:1987ia,Gross:1987fk}.
Consider for example the identity element $|I\ra$  defined as
\be
|I\ra =b_+(\hp)b_-(\hp) |\tilde{I}\ra,
\ee
where
\be
|\tilde{I} \ra= \exp \left( -\frac12 \sum_{n\ge 0} (-1)^n  a_n^{\dagger}
a_n^{\dagger}+ \sum_{n\ge 1} c_{-n} b_{-n})\right) c_0 c_1 |0 \ra \ ,
\ee
and $| 0 \rangle$ is the vacuum invariant with respect to $SL(2,R)$
subgroup of the Virasoro algebra.  $|\tilde I \rangle$ satisfies the
relations
\beq\label{p}
&& (a_m + (-1)^m a_{-m}) |\tilde{I} \rangle =  0, \nonumber \\
&& (c_m + (-1)^m c_{-m})| \tilde{I} \rangle =  0, \\
&& (b_m - (-1)^m b_{-m})| \tilde{I} \rangle =  0 \ \nonumber ,
\eeq
whereas $|I \rangle$ violates (\ref{p}) for the $c$ fields due to the
presence of extra factors $b_+(\hp)b_-(\hp)$.  With these extra factor
of $b$ fields, $|I \rangle$ is BRST invariant
\begin{equation}
Q |I \rangle = 0 \ .
\end{equation}
Using $\langle I|$, we can evaluate
\begin{equation}\label{jj}
\int V A = \langle I | V(\hp) | A \rangle, \label{op1}
\end{equation}
explicitly in terms of the oscillators.  Formal argument for gauge
invariance at the linearized level follows immediately from the fact
that $[Q,V]=0$ and $\langle I|Q = 0$. There are some subtleties,
however, that needs to be addressed before this expression can be made
completely well defined.  To be concrete, let us take $V$ to be the
vertex operator of the closed string tachyon
\begin{equation}
V(\sigma_+,\sigma_-) = c_+ (\sigma_+) c_- (\sigma_-) \norm{e^{i 
 k X_+(\sigma_+) /2}}  \,  \norm{e^{i 
 k X_- (\sigma_-)/2}}   . \label{tachyon}
\end{equation}
One of the subtleties arises from the fact that the insertion of
$c_+(\hp) c_-(\hp)$ of $V$ and $b_+(\hp) b_-(\hp)$ of $\langle I |$
collides on the world sheet.  One way to avoid this subtlety is to
simply interpret
\begin{equation}
\langle \tilde I | b_+(\hp) b_-(\hp) c_+(\hp) c_-(\hp) {\cal O}(\hp)
= {\cal N} \langle \tilde I |  {\cal O}(\hp),
\end{equation}
where
$
{\cal N}  = \delta^2(0),
$
is an infinite factor obtained from evaluating the commutator of
conjugate fields at the same point. Since this is just an overall
multiplicative factor,  we might as well not include it in our
definition of the gauge invariant operator. In other words, we 
define a quantity of the form
\be\label{jj1} O = \langle \tilde I |{\cal O}(\hp)
|A \ra \ . \label{op2} \ee
This quantity is well defined and can be evaluated for any given closed
string vertex operator.  For example, using the closed string tachyon,
this evaluates to
\begin{eqnarray}
O_T& =&\langle \tilde I | e^{i k X(\pi/2)/2} |A \rangle   \label{Otac}\\
& =& \langle 0|c_0 c_{-1} \exp\left(  -\frac12 \sum_{n\ge 0} (-1)^n  a_n
a_n+ \sum_{n\ge 1} c_{n} b_{n} 
- \sqrt{2 \alpha'} k \,  \left({a_0\over 2} + \sum_{n \ge 1}  
 \sqrt{{1 \over n}}(-1)^n a_{2n} \right)
\right) | A \rangle \nonumber
\end{eqnarray}
where $X = X_+ + X_-$.  This
expression is in a complete agreement with the open-closed interaction
vertex of Shapiro and Thorn \cite{Shapiro:1987gq,Shapiro:1987ac} for
the closed string tachyon.

We need to verify that this quantity is invariant with respect to
gauge transformation at the linear and the non-linear level.  At the
linear level, we need to verify that
\begin{equation}
Q \, e^{i k X(\pi/2)/2} |\tilde I \rangle = 
Q\,  \exp\left(\sqrt{2 \alpha'} k \left({a_0\over 2} + \sum_{n \ge 1}   \sqrt{{1 \over n}}(-1)^n a_{2n} \right)\right) | \tilde I \rangle = 0.
\end{equation}
This follows from the fact that \cite{Gross:1987fk}
\be
Q | \tilde I \rangle  = -16  \sum_{n\ge 1} n(-1)^n c_{2n} | \tilde I \rangle,
\ee
and
\beq
&& [Q, \exp\left(\sqrt{2 \alpha'} k \left({a_0\over 2} + \sum_{n \ge 1}  
 \sqrt{{1 \over n}}(-1)^n a_{2n} \right)\right)]| \tilde I \rangle =\nonumber \\ 
&& \exp\left(\sqrt{2 \alpha'} k \left({a_0\over 2} + \sum_{n \ge 1}  
 \sqrt{{1 \over n}}(-1)^n a_{2n} \right)\right)
2 (2 \alpha' k^2)
 \sum_{n\ge 1} n(-1)^n c_{2n} | \tilde I \rangle \label{normalorder},
\eeq
so that they vanish for
\begin{equation}
2 \alpha' k^2 = 8
\end{equation}
which is precisely the on-shell condition for the closed string tachyon.
This establishes the gauge invariance of (\ref{op2}) at the
linear level.

There is a rather subtle point here.  One
could very easily  reach a different conclusion by performing
the calculation in a  different order. Namely
\beq
&&Q\,  e^{ikX(\pi/2)/2} |\tilde I \rangle 
 = [Q,e^{ikX(\pi/2)/2}]|\tilde I \rangle + e^{ikX(\pi/2)/2} Q | \tilde I \rangle \\
&&\qquad  = (2 \alpha' k^2 - 16)  \sum_{n\ge 1} n(-1)^n c_{2n} | \tilde I
 \rangle. \nonumber 
\eeq
This leads to the condition $2 \alpha' k^2 = 16$ for BRST invariance
which is off by a factor of two compared to the on-shell condition of
the closed string tachyons.  The origin of this discrepancy can be
traced to an anomaly in associativity\footnote{Associativity anomaly
in string field theory was originally formulated
in \cite{Horowitz:1987yz}.}
\begin{equation}
Q ( e^{i k X(\pi/2)/2} |\tilde I \rangle) \neq  (Q e^{i k X(\pi/2)/2}) | \tilde I \rangle \label{anomaly} \ .
\end{equation}
Since we are interested in understanding the BRST invariance of a
state, $ |T,p \rangle = e^{i k X(\pi/2)/2} |\tilde I \rangle, $ we should
show that the left hand side, and not the right hand side, of
(\ref{anomaly}) is zero. Computing $e^{ikX(\pi/2)} |\tilde I\rangle$ {\em
before} acting with $Q$ essentially amounts to normal ordering this
expression so that $Q e^{ikX(\pi/2)/2} | \tilde I \rangle$ becomes
(\ref{normalorder}).  Put differently, understanding that (\ref{jj})
or (\ref{jj1}) should be normal ordered defines $|T,p \rangle$
unambiguously, and the condition for BRST invariance of $|T,p \rangle$
becomes the on-shell condition of the closed string tachyon.

The other main subtleties are in the proof of gauge invariance at the
non-linear level. This is equivalent to showing
\be
\langle \tilde I| {\cal O}  |A* \Lambda \rangle -
\langle \tilde I| {\cal O} |\Lambda *A  \rangle =0 \ , 
\ee
or more explicitly,
\be
{}_{123}\la V_3 |{\cal O}| \tilde{I} \ra_3 |A\ra_1 |\Lambda\ra_2 -
{}_{123}\la V_3 |{\cal O}| \tilde{I} \ra_3 |A\ra_2 |\Lambda\ra_1 = 0 .
\ee
Where $|V_3\rangle$ is the three string
vertex whose details will be given shortly.
The subtleties arose from the ambiguity in inserting an operator at
the midpoint which is precisely the point where the local coordinate
of the glued world sheet of the form illustrated in figure \ref{figb}
is singular. One can again regulate this potential singularity by
``point splitting''
\be O = \lim_{\epsilon=0} \langle \tilde I |{\cal O}(\hp-\epsilon)
|A \ra. \label{op3} \ee
For this quantity, establishing the gauge invariance at the non-linear
level amounts to showing that
\be\label{8b}\lim_{\epsilon\rightarrow 0} \;
 _{123} \la V_3|{\cal O}(\hp-\epsilon)
| \tilde{I}\ra_{3},
\ee
is symmetric in $1$ and $2$.  For technical reasons that will be
explained later, it is more convenient to show an equivalent statement
that
\be\lim_{\epsilon\rightarrow 0}\;_3  \la \tilde{I} |{\cal O}(\hp-\epsilon)
 | V_3\ra_{123},
\ee
is anti-symmetric in $1$ and $2$.\footnote{To see this note
that $$_{123}\langle V_3 |{\cal O} |\tilde I \rangle_3 =
{}_{14}\langle V_2|\,  {}_{25}\langle V_2| \,
{}_3 \langle \tilde I_3 | {\cal O} |V_3 \rangle_{345}.$$
Thus under the exchange $1 \leftrightarrow 2$,
$$_{132}\langle V_3 |{\cal O} |\tilde I \rangle_3 =
{}_{24}\langle V_2|\,  {}_{15}\langle V_2| \,
{}_3 \langle \tilde I_3 | {\cal O} |V_3 \rangle_{345} \ .
$$
Combining this with  \cite{Kishimoto:2001ac}
$$\langle V_2| = {} _{12}(\langle 0 | c_{-1}) \exp \left[ - a_n^1 C_{nm}
 a_m^2 - (c_n^1 C_{nm} b^2_m + c_n^2 C_{nm} b^1_m) \right] (c_0^1 + c_0^2), $$
implies that
$${}_{24}\langle V_2|\,  {}_{15}\langle V_2| \,
{}_3 \langle \tilde I | {\cal O} |V_3 \rangle_{345}
=
-  {}_{15}\langle V_2| \, {}_{24}\langle V_2|\,
{}_3 \langle \tilde I | {\cal O} |V_3 \rangle_{345}
=
-  {}_{14}\langle V_2| \, {}_{25}\langle V_2|\,
{}_3 \langle \tilde I | {\cal O} |V_3 \rangle_{354} \ .
$$
Therefore, if $_{123} \langle V_3 | {\cal O} | \tilde I \rangle_3$ is
even under $1 \leftrightarrow 2$, then $_3 \langle \tilde I | {\cal O}
| V_3 \rangle_{123}$ must be odd.  }
The way we prove this is by showing that the matter part is symmetric
while the ghost part is anti-symmetric.

Let us start with the matter part.  For simplicity we consider the
closed string tachyon.  Using the mode expansion and eq.(\ref{p}),  one
finds
\begin{equation}
 \langle \tilde{I} |e^{i  k X(\sigma)/2}
= \langle \tilde{I} | 
\exp\left[{i \over 2}  \, \sqrt{2 \alpha'} k \, \beta_n(\sigma)
(1 + C)_{nm} a_m\right] ,
\end{equation}
where
\begin{equation}
\beta_0(\sigma) = {i  \over 2}, \qquad \beta_m(\sigma)
= i  \sqrt{{1 \over m}} \cos(m \sigma),  \,
\end{equation}
and $C$ is defined like in \cite{Rastelli:2001jb}
\be C_{nm}=\delta_{nm} (-)^n.  \ee 
Without loss of generality we can assume that $\sigma\leq \hp$.  Then,
since $C_{nm} \beta_m(\sigma) = \beta_n(\pi - \sigma)$, we have
\begin{equation}\label{7}
\langle \tilde{I} |e^{i  k X(\sigma)/2}
= \langle \tilde{I} | \exp\left[{i \over 2} \, \sqrt{ 2 \alpha'} k \, (\beta_n(\sigma) +
 \beta_n(\pi - \sigma)) a_m\right] .
\end{equation}
To compute (\ref{8b}) one uses the definition of the three-string 
vertex
\be
|V_3 \rangle = \exp \left[ - \sum_{n,m \ge 0} {1 \over 2}
a^r_{-n} V^{rs}_{nm} a^s_{-m} \right] | 0 \rangle.
\ee
Here $n$ and $m$ run over the stringy modes while $r$ and $s$ specify
the string field state. That is, $r,s=1,2,3$.  We would like to study
the behavior under $1 \leftrightarrow 2$. Thus it is convenient to
separate the $3$ sector
\be
|V_3 \rangle = \exp \left[ {1 \over 2} (a^{\dag 1} \ a^{\dag 2} )
\left(\begin{array}{cc} V_{11} & V_{12} \\ V_{21} & V_{22} \end{array}\right)
\left(\begin{array}{c} a^{\dag 1} \\  a^{\dag 2} \end{array}\right)  -
(a^{\dag 1} \ a^{\dag 2})
\left(\begin{array}{c} V_{13} \\ V_{23} \end{array}\right) a^{\dag 3}
- {1 \over 2} a^{\dag 3} V_{33} a^{\dag 3} \right] | 0 \rangle_{123}\ . \nonumber
\ee
With the help of standard relation
\beq
&&\la 0|\exp (\lambda_i a_i -\frac12 P_{ij} a_i a_j) \exp(\mu_i a_i^{\dagger}
-\frac12 Q_{ij} a_i^{\dagger} a_j^{\dagger} )|0 \ra \nonumber \\
&&=\det(K)^{-\frac12}\exp(\mu K^{-1} \lambda -\frac12 \lambda Q K^{-1}
\lambda -\frac12 \mu K^{-1} P \mu), \;\;\;\;K=1-PQ \ ,
\eeq
and using some identities \cite{Rastelli:2001rj}
\beq
&& M_{rs} = C V_{rs}, \nonumber \\
&& M_{11}+M_{12}+M_{21}=1, \qquad
M_{12} M_{21}= M_{11}^2- M_{11}, \\
&& M_{12}^2+M_{21}^2 =1 - M_{11}^2, \qquad
M_{12}^3+M_{21}^3=(1-M_{11})^2(1+2M_{11})\ ,\nonumber
\eeq
we find\footnote{We have dropped the overall determinant factor of
$\det(1 - M_{11})^{-1}$ since it does not affect the proof of gauge
invariance.}
\begin{eqnarray}
&& \lefteqn{ _3 \langle \tilde{I}
 |e^{i k X(\sigma)/2} | V_3 \rangle_{123}}
\nonumber \\
&&  = \exp\left[ - (a^{\dag 1} \ a^{\dag 2})
\left\{  \left(\begin{array}{c}1 \\ 1 \end{array}\right) -
\left(\begin{array}{c}1 \\ -1 \end{array}\right) {M_{12} - M_{21} \over 1-
M_{11}} \right\}
{1 \over 2}\, \sqrt{2 \alpha'} k \,  (\beta(\sigma) + \beta(\pi-\sigma)) \right. \nonumber \\
&&\left. - {1 \over 2} (a^{\dag 1}\ a^{\dag 2})
\left(\begin{array}{cc} 0 & C \\ C & 0 \end{array}\right)
\left(\begin{array}{c} a^{\dag 1} \\ a^{\dag 2} \end{array}\right) \right]| 0  \rangle_{12}.
\end{eqnarray}
After some manipulations which are explained in the appendix, one
finds that
\begin{equation}\label{7a}
{M_{12} - M_{21} \over 1 - M_{11}} (\beta(\sigma) + \beta(\pi-\sigma)) = \beta(\sigma) - \beta(\pi-\sigma), \qquad \sigma < \hp \ .\end{equation}
Therefore,
\be
_3 \langle \tilde{I}
|e^{i k X(\sigma)/2} | V_3 \rangle_{123}
=  \exp\left[ - {1 \over 2} \, \sqrt{2 \alpha'} \, k (a^{\dag 1} \beta(\sigma) 
+  \ a^{\dag 2} \beta(\pi-\sigma))    - 
a^{\dag 1} C  a^{\dag 2} \right] | 0 \rangle_{12}\ ,\nonumber
\ee
which is indeed symmetric in $1$ and $2$ in the limit $\sigma\rightarrow\hp$.

The calculation for the ghost part goes along similar lines.
We would like to compute
\begin{equation}
_3 \langle \tilde I | V_3 \rangle_{123},
\end{equation}
and show that it is antisymmetric with respect to the exchange $1
\leftrightarrow 2$. We  begin with the  explicit expressions for
$\langle \tilde I |$ and $|V_3 \rangle$ in terms of oscillators
\cite{Gross:1987fk,Kishimoto:2001ac}
\begin{eqnarray}
&&  \langle \tilde I | = \langle V_2 | \tilde I \rangle = \langle 0 |
c_{-1} c_0
\exp[- \!\! \sum_{n,m \ge 1} c_{n} C_{nm} b_{m}], \nonumber \\
&& | V_3 \rangle_{123}   =
 \exp[ -\!\! \sum_{{n \ge 1 \atop \ m \ge 0}} c_{-n}^r \tv^{rs}_{nm}
 b^s_{-m} ] (c_0 c_1 | 0 \rangle)_{123} \ .
\end{eqnarray}

In order to contract along $b^3$ and $c^3$, we  separate the 3
sector as we did in the matter part.  We  also analyze the zero
mode and the non-zero mode parts of $|V_3 \rangle$ separately.
\begin{eqnarray}
|V_3 \rangle & = &
\exp \left[\rule{0ex}{5.5ex} - \sum_{n,m \ge 1} \left\{
c_{-n}^3 \tv{}^{33}_{nm} b_{-m}
+ c_{-n}^3 (\tv{}_{nm}^{31} \ \tv{}_{nm}^{32}) \left( \begin{array}{c} b^1_{-m} \\ b^2_{-m} \end{array}\right) \right.\right. \nonumber \\
&& \left. \qquad
+ (c_{-n}^1\ c_{-n}^2) \left(\begin{array}{c} \tv{}_{nm}^{13} \\ \tv{}_{nm}^{23}
\end{array}\right) b_{-m}^3
+ (c_{-n}^1\ c_{-n}^2) \left(\begin{array}{cc}\tv{}_{nm}^{11} & \tv{}_{nm}^{12}
\\ \tv{}_{nm}^{21} & \tv{}_{nm}^{22} \end{array} \right) \left(\begin{array}{c} b^1_{-m} \\ b^2_{-m} \end{array}\right) \right\} \nonumber \\
&& \qquad \left. - \sum_{n \ge 1}
\left\{ c^3_{-n} (\tv{}^{31}_{n0}\ \tv{}^{32}_{n0}\ \tv{}^{33}_{n0})
\left(\begin{array}{c} b^1_0 \\b^2_0 \\b^3_0\end{array}\right) \right.\right. \\
&& \qquad \left. \left.
+ (c^1_{-n}\ c^2_{-n}) \left( \begin{array}{ccc}
\tv{}^{11}_{n0} & \tv{}^{12}_{n0} & \tv{}^{13}_{n0}   \nonumber \\
\tv{}^{11}_{n0} & \tv{}^{12}_{n0} & \tv{}^{13}_{n0}   \end{array}\right)
\left(\begin{array}{c} b^1_0 \\b^2_0 \\b^3_0\end{array}\right)
\right\} \right] (c_0 c_1 | 0 \rangle)_{123} \ .
\end{eqnarray}
We can now compute the contraction over the non-zero modes of $b^3$
and $c^3$ using the relation given in \cite{Kishimoto:2001ac}
\begin{eqnarray}
&&\langle 0 | c_{-1} c_0   \exp\left[-c_n P_{nm} b_{m}\right]
\exp\left[-c_{-n} Q_{nm} b_{-m} - \lambda_n b_{-n} - c_{-n} \mu_n \right]
c_0 c_1|  0 \rangle \nonumber \\
&&\qquad\qquad\qquad=
\det(1 + PQ) \langle 0 | c_{-1} c_0  \exp[\lambda (1+PQ)^{-1} P \mu] c_0 c_1|  0 \rangle,
\end{eqnarray}
where
\begin{eqnarray}
&& M  =  C, \;\;\;\;
N =  \tv{}^{33}, \;\;\;\;
\lambda  =
(c_{-n}^1 \ c_{-n}^2) \left(\begin{array}{c} \tv{}_{nm}^{13} \\ \tv{}_{nm}^{23}
\end{array}\right), \nonumber \\
&& \mu  =
 (\tv{}_{nm}^{31}\ \tv{}_{nm}^{32}) \left( \begin{array}{c} b^1_{-m} \\ b^2_{-m} \end{array}\right) +
(\tv{}^{31}_{n0} \ \tv{}^{32}_{n0} \ \tv{}^{33}_{n0})
\left(\begin{array}{c} b^1_0 \\b^2_0 \\b^3_0\end{array}\right) \ .
\end{eqnarray}
Now, using the variables
\begin{equation}\tilde M^{rs} = - C \tv{}^{rs},
\end{equation}
and the relations \cite{Kishimoto:2001ac}
\beq
&& \tilde M_{11}+\tilde M_{12}+\tilde M_{21}=1,  \qquad
\tilde M_{12} \tilde M_{21}= \tilde M_{11}^2- \tilde M_{11}, \nonumber \\
&& \tilde M_{12}^2+\tilde M_{21}^2 =1 - \tilde M_{11}^2, \qquad
\tilde M_{12}^3+\tilde M_{21}^3=(1-\tilde M_{11})^2(1+2 \tilde M_{11}),\\
&& \tv^1{}^s_0 + \tv^2{}^s_0+ \tv^2{}^s_0 = 0, \qquad
\tv^r{}^1_0 + \tv^r{}^2_0 + \tv^r{}^3_0 = 0,\nonumber \\
&& \tv^2{}^1_0 = - {\tilde M_{12} \over 1 - \tilde M_{11}} \tv^1{}^1_0,
\qquad
\tv^1{}^2_0 = - {\tilde M_{21} \over 1 - \tilde M_{11}} \tv^1{}^1_0 , \nonumber
\eeq
one can show that
\begin{equation}
{}_3 \langle \tilde I | V_3 \rangle_{123}=
{}_3 \langle 0| c^3_{-1} c^3_0 \exp\left[
c^1 C b^2 + c^2 C b^1
+ (c^1 - c^2) \Omega (b^1_0 - b^2_0) + (c^1 - c^2) \Gamma b^3_0
 \right] (c_0 c_1 |0 \rangle)_{123}  \nonumber 
\end{equation}
where
\begin{eqnarray}
\Omega &=& - {1 \over 1 - \tilde{M}_{11}} \tv^1{}^1_0 ,\nonumber \\
\Gamma & = & (\tilde V^1{}^2_0 - \tilde V^2{}^1_0) \ .
\end{eqnarray}
Following the notation of \cite{Kishimoto:2001ac}, we have suppressed
the subscript $n$ and $m$ which take on integer values greater than or
equal to one in expressions like in expressions like $c^r_n$, $b^r_n$,
$\tilde V{}_{nm}^{rs}$ and $\tilde V{}_{n0}^{rs}$.  Finally,
recalling that $\langle 0 | c_{-1} c_0 c_1 | 0 \rangle = 1$, we arrive
at
\begin{equation}
{}_3 \langle  \tilde I | V_3 \rangle_{123}
=
(c^1 - c^2) \Gamma  \exp\left[ 
c^1 C b^2 + b^2 C b^1 
+ (c^1 - c^2) \Omega (b^1_0 - b^2_0) \right]
(c_0 c_1 |0 \ra)_{123}  ,
\end{equation}
which is clearly odd under the exchange $1 \leftrightarrow 2$.

This rigorously establishes the gauge invariance of (\ref{op2}) when
${\cal O}$ is the closed string tachyon.  It would be interesting to
extend this calculation  to other closed string modes.

\section{Vacuum string field theory}

In this paper we focused on Witten's open string field theory.
Recently, some compelling numerical evidence
\cite{Sen:1999nx,Moeller:2000xv,Ellwood:2001py} that there exists a
non-trivial saddle point in this theory was presented.  This saddle
point has a non-vanishing tachyon expectation value and is conjectured
to describe the vacuum where the D25-brane have decayed away.
However, the description of the fluctuation around this new saddle
point in terms of Witten's original string fields is very complicated
and to date have been explored only using numerical techniques
\cite{Ellwood:2001py}.  To overcome this challenge, a new theory,
known as vacuum string field theory \cite{Rastelli:2000hv}, was
conjectured to describe the fluctuations about the non-trivial vacuum
of Witten's original theory.  We close this paper by commenting on the
nature of gauge invariant observables for this theory.\footnote{Gauge
invariant observables of vacuum string field theory was also
considered in \cite{Gross:2001yk}. The relation between these
operators and the operators described in this paper is not clear.}

The action of vacuum string field theory is given by
\begin{equation}\label{7mm}
S = \int A * {\cal Q} A + {2 \over 3} A*A*A \ , \label{vsft}  
\end{equation}
where ${\cal Q}$ consists only of ghosts
\begin{equation}\label{4a}
{\cal Q} = a_0 c_0 + \sum_{n=1} a_n (c_n + (-1)^n c_n)  \ ,
\end{equation}
and $a_n$ are in general arbitrary numerical coefficients.  Because the
operator ${\cal Q}$ has trivial cohomology, there are no perturbative
open string states in this theory, justifying the conjecture.

If the conjecture of \cite{Rastelli:2000hv} is correct and the vacuum
string field theory is indeed equivalent to Witten's theory around the
shifted vacuum, we expect to find the same set of gauge invariant
operators in vacuum string field theory as we did in Witten's
theory. After all, these operators correspond to closed strings which
continue to exist in the absence of D-branes. Let us therefore examine
how the argument for gauge invariance of (\ref{op2}) is modified when
the action is (\ref{7mm}).

In vacuum string field theory, gauge transformation acts on the string
fields according to
\begin{equation}
\delta A = {\cal Q} \Lambda + [A,\Lambda] \ .
\end{equation}
Consider an expression of the form (\ref{op2}) considered in the
previous section. Just as in the previous section, we are interested
in whether (\ref{op2}) is invariant with respect to the linear and the
non-linear contributions to the gauge transformations.  Since the
vacuum string field theory differs only in the choice of ${\cal Q}$,
the only possible difference from the analysis of the previous section
will arise from the effect of the linear term.

In Witten's cubic theory, gauge invariance of (\ref{op2}) at the
linear level required the BRST operator to commute with the closed
string vertex. This constrained the momentum of (\ref{op2}) to lie on
the mass shell of the corresponding closed string. In the case of the
vacuum string field theory, however, this constraint is much simpler
since ${\cal Q}$ acts trivially on the matter sector. Consider for
example the tachyon vertex operator (\ref{tachyon}). Since
(\ref{tachyon}) does not contain any dependence on the $b$ field, it
commutes with ${\cal Q}$ trivially. We therefore see that there are
more gauge invariant operators in vacuum string field theory.

The fact that the on-shell condition is relaxed in the vacuum string
field theory giving rise to a mismatch in the spectrum of gauge
invariant observables appears to be a generic feature. The structure
of gauge transformations involving ${\cal Q}$ made purely out of
ghosts is simply not restrictive enough. This mismatch appears to
point to the conclusion that the conjecture of \cite{Rastelli:2000hv}
is at best singular.  In order for ${\cal Q}$ to have the possibility
of imposing mass shell condition on the (\ref{op2}), ${\cal Q}$ must
depend at some degree on matter fields. The challenge is to find
${\cal Q}$ with trivial cohomology that imposes the correct on-shell
condition on the gauge invariant operators.  One way to approach this
difficulty is \cite{barton} to regulate ${\cal Q}$ with an explicit
dependence on the matter part.  It would be very interesting to see if
such a prescription does in fact give rise to the correct physics.

\section*{Acknowledgements}

We thank S. Giddings, D. Gross, L. Rastelli, A. Sen, and B. Zwiebach for
discussions.  This work is supported in part by DOE grant
DE-FG02-90ER40542. The work of AH is also supported in part by the the
Marvin L.~Goldberger fellowship.

\section*{Appendix A: Proof of eq.(\ref{7a})}
         {\setcounter{section}{1} \gdef\thesection{\Alph{section}}}
          {\setcounter{equation}{0}}

To prove eq.(\ref{7a}) we start with some relations found in
\cite{Gross:1987ia}
\beq
&& V_{11}=\frac13 (C+U+\bar{U}),\nonumber \\
&& V_{12}=\frac16 (2C -U -\bar{U}) +\frac16 i\sqrt{3} (U-\bar{U}),\\
&& V_{21}=\frac16 (2C-U-\bar{U}) -\frac16 i\sqrt{3} (U-\bar{U}).\nonumber
\eeq
Where $\bar{U}=CUC$ and $U$ satisfy
\be\label{7y}
(1-Y)E(1+U)=0,\;\;\;(1+Y)\frac1E (1-U)=0,\;\;\;Y=-\frac12C+\frac12\sqrt{3}X.
\ee
With
\be\label{7q}
(E^{-1})_{nm}=\delta_{nm}\sqrt{\frac12 n}+\delta_{n0}\delta_{m0},
\ee
and $X_{nm}$ are the Fourier components of the operator
\be
X(\sigma, \sigma')=i\left( \Theta(\hp -\sigma)-\Theta(\sigma -\hp) \right)
\delta(\sigma+\sigma'-\pi),
\ee
which can be written explicitly as
\beq\label{7v}
&& X_{0m}=\frac{i\sqrt{2}}{\pi m}(1-(-1)^m)(-1)^{(m-1)/2},\nonumber \\
&& X_{nm}=\frac{i}{\pi} (-1)^{(n-m-1)/2}
(1-(-1)^{n+m})\left(\frac{1}{n+m}
+\frac{(-1)^m}{n-m}\right).
\eeq
When decompose these matrices into two by two block matrices associated
with odd and even indices they take the form \cite{Gross:2001rk}
\beq\label{7i}
C= \left(\begin{array}{cc} -1 & 0 \\ 0 & 1 \end{array}\right),\;\;
X=\left(\begin{array}{cc} 0 & X_{oe} \\ X_{eo} & 0 \end{array}\right),\;\;
U=\left(\begin{array}{cc} U_{oo} & U_{oe} \\ U_{eo} & U_{ee}
 \end{array}\right),\;\;
\bar{U}=\left(\begin{array}{cc} U_{oo} & -U_{oe} \\ -U_{eo} & U_{ee}
 \end{array}\right).
\eeq
From the $oo$ and $oe$ components of (\ref{7y}) we have
\cite{Gross:2001rk}
\beq\label{7h}
&& (1+U_{oo})-i\sqrt{3} F U_{oe}=0,\;\;\; -i\sqrt{3}F(1+U_{ee})+U_{oe}=0,
\nonumber \\
&& 3(1-U_{oo}) -i\sqrt{3}(F^T)^{-1} U_{eo}=0,\;\;\;
i\sqrt{3}(F^T)^{-1} (1-U_{ee})-3U_{oe}=0.
\eeq
where\footnote{Note that we use $F$ to denote what was defined as $M$
in \cite{Gross:2001rk} to avoid confusion with $M_{11}$, $M_{12}$ and
$M_{21}$.}

\be
F=E^{-1} X_{oe} E, \;\;\;\;(F^T)^{-1}=EX_{oe} E^{-1}.
\ee
From (\ref{7i}) it follows that
\beq
 M_{12}-M_{21}=\frac{2i}{\sqrt{3}}
\left(\begin{array}{cc} 0 & -U_{oe} \\ U_{eo} & 0 \end{array}\right),\;\;\;
 1-M_{11}=\frac23\left(\begin{array}{cc} 1+U_{oo} & 0 \\ 0 & 1-U_{ee}
\end{array}\right).
\eeq
Combining this with (\ref{7h}) we get
\be
U_{oe} = {i \over \sqrt{3}} (F^T)^{-1} (1 - U_{ee}),\;\;\;\;
U_{eo} = - {i \over \sqrt{3}} F^{-1} (1 + U_{oo}).
\ee
Therefore,
\begin{equation}
M_{12} - M_{21} = - {2 \over 3}
\left( \begin{array}{cc} 0 & (F^{T})^{-1} \\ F^{-1} & 0\end{array}\right)
\left( \begin{array}{cc} 1 - U_{ee} & 0\\0  & 1 + U_{oo} \end{array}\right).
\end{equation}
And so
\begin{equation}\label{7p}
{M_{12} - M_{21} \over 1 - M_{11}}
= \left( \begin{array}{cc} 0 & E X_{oe} E^{-1} \\
E^{-1} X_{eo} E & 0 \end{array}\right).
\end{equation}
Using the fact that $\beta_n(\sigma) + \beta_n(\pi-\sigma)$ is
non-vanishing only for even $n$, we can write
\be
{M_{12} - M_{21} \over 1 - M_{11}} (\beta(\sigma) + \beta(\pi-\sigma)) = 
E X_{oe} E^{-1} (\beta(\sigma) + \beta(\pi-\sigma)) \ .
\ee
Now using (\ref{7q}) and  (\ref{7v}), we get eq.(\ref{7a}).

\bibliography{sft}\bibliographystyle{utphys}

\end{document}